# The Tiny Tera: A Packet Switch Core

**Nick McKeown, Martin Izzard\*,
Adisak Mekkittikul, William Ellersick, Mark Horowitz**

Departments of Electrical Engineering and Computer Science

Stanford University, Stanford, CA 94305-4070

\*Communications Laboratory

DSP R&D Center, Corporate Research & Development

Texas Instruments, Incorporated, PO Box 655474, MS446, Dallas, TX 75265

*Abstract* — **In this paper, we present the *Tiny Tera*: a small packet switch with an aggregate bandwidth of 320Gb/s. The *Tiny Tera* is a CMOS-based input-queued, fixed-size packet switch suitable for a wide range of applications such as a high-performance ATM switch, the core of an Internet router or as a fast multiprocessor interconnect. Using off-the-shelf technology, we plan to demonstrate that a very high-bandwidth switch can be built without the need for esoteric optical switching technology. By employing novel scheduling algorithms for both unicast and multicast traffic, the switch will have a maximum throughput close to 100%. Using novel high-speed chip-to-chip serial link technology, we plan to reduce the physical size and complexity of the switch, as well as the system pin-count.**

## 1 Introduction

The case for high-performance networking is overwhelming; the traffic on the Internet continues to grow at over 30% per month with no apparent end in sight. As a result, recent years have witnessed an increasing interest in high-speed networks supporting IP or ATM. High speed networks need high performance routers and switches. Recently, there has been a blurring of the line between switches and routers in attempts to combine the benefits of both, but this alone will not satisfy the increased bandwidth requirements. We believe it is necessary to ensure that the bandwidth bottleneck in network elements is fundamental, rooted in memory or interconnect bandwidth limitations.

In an attempt to provide industry with a novel switching element, we are developing and building the *Tiny Tera*: a small, high-bandwidth, single-stage switch. The *Tiny Tera* has 32 ports, each operating at 10Gb/s (approximately the OC-192 rate), switching fixed-size packets. The switch distinguishes four classes of traffic, and includes efficient support for multicast. We aim to demonstrate that it is possible to build a compact switch with an aggregate bandwidth of some 320Gb/s using currently available CMOS technology, with a central hub no larger than a soda can. Such a switch could serve as a core for applications as diverse as an ATM Switch or an Internet Router.

The *Tiny Tera* is an input-buffered switch allowing it to be the highest bandwidth switch possible given a particular CMOS and Memory technology. The switch consists of three logical elements: Ports, a central Crossbar switch, and a central Scheduler. Packets are queued at the port on entry to the switch and optionally prior to exit. The scheduler has a map of all the ports' queue occupancy and decides the crossbar configuration every packet-time.

---

1. This work is supported by Texas Instruments, Inc. and the Center for Telecommunications at Stanford University.

Input-queueing, parallelism, and tight integration provide the keys to such a high-bandwidth switch. Input-queueing reduces the memory bandwidth requirements; when packets are queued at the input, the buffer memories need run no faster than the line rate, so there is no need for the speedup that is required in output-queued switches. The long-standing view has been that packets in an input-queued switch suffer poor performance due to *head of line* (HOL) blocking [2], but we have developed novel scheduling algorithms to reduce the effects of HOL blocking for both unicast and multicast traffic. For unicast traffic, we use a well-known buffering scheme called Virtual Output Queueing (VOQ) [1] in which each input maintains a separate queue for each output. Motivated by DEC's PIM algorithm [3], we use a novel, fast, fair, and efficient scheduling algorithm called *i*SLIP, that achieves a throughput close to 100% [7] [8], yet is able to make a scheduling decision in less than 40ns in current technologies. For multicast traffic, we are developing algorithms based on fanout splitting and residue concentration ideas[4].

In order to realize a compact switch with such high aggregate bandwidth, the *Tiny Tera* interconnect must be optimized to support its massive bandwidth yet allow the use of connectors, variable length physical paths, and ICs with reasonable pin-counts and power-consumption. We are developing a Serial Link circuit block to provide this IC-to-IC physical link. The VLSI engines will see the link as a byte interface, yet the interconnect will be a single low-swing differential pair. The block contains: MUX; DMUX; Clock Multiplication PLL; Clock Recovery PLL; Line Drivers; Line Receivers.

# 2 Switch Architecture

As shown in Figure 1, the *Tiny Tera* switch consists of three main parts: a parallel sliced self-routing crossbar switch, a centralized scheduler for configuring the crossbar, and 32 ports, each operating at 10Gb/s. When a packet arrives at a port, it is buffered in an input queue according to its destination, priority class, and whether it has a single destination (unicast) or multiple destinations (multicast). The packet awaits a decision by the scheduler allowing it to traverse the crossbar switch fabric. At the beginning of each fixed-length packet time (which we will call a *slot*), the scheduler examines the contents of all input queues, decides upon the configuration of the crossbar, and chooses a set of conflict-free connections between inputs and outputs. The scheduling decision is passed back to the ports which communicate the configuration information to the crossbar slices, and then transmit packets into the crossbar. Packets leaving the crossbar are buffered in the output queues where they await transmission to the external line.

## 2.1 Scheduler and Crossbar Switch Hub

The scheduler is situated in the central hub (Figure 1b) and is connected to all ports to allow it to easily gather queue status updates and issue configuration data. It is the switch pipeline root, because it sources configurations (grant-tokens) that trigger the release of back pressure mechanisms throughout the switch. This allows for a flexible pipeline architecture. The scheduler implements algorithms that provide efficient use of the crossbar bandwidth for both unicast and multicast.

The sliced crossbar switch makes up the rest of the central hub (Figure 1b). Each slice of the *Tiny Tera* is a printed circuit board containing a 1-bit $32 \times 32$ crossbar chip.

The crossbars and the scheduler are connected to each port using high-speed serial links operating at multiple Gb/s. The links are described in more detail in Section 4.

The advantages of this central sliced crossbar switch hub are:

- The crossbar slice is extremely simple. No traces need to cross.

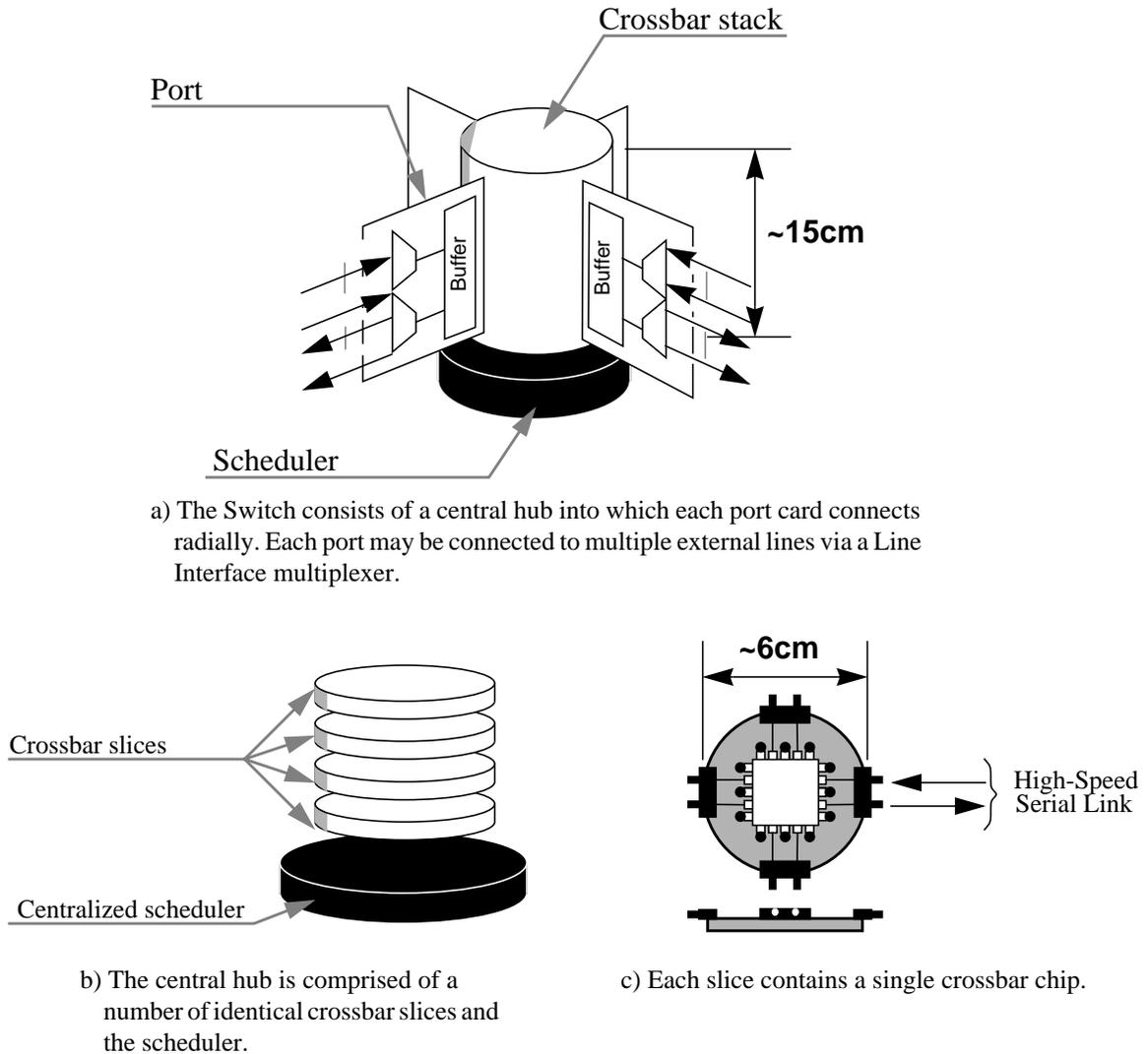

Figure 1  Architecture and detail of the *Tiny Tera* switch (4 port example).

- The trace lengths connecting each port to the crossbar are all of similar and minimum length. This reduces skew and crosstalk. It also means that each slice can be very small; in the *Tiny Tera*, each slice will be approximately 6cm in diameter.
- Extremely high aggregate bandwidths are achievable by switching multiple bits in parallel; as a result, the switch can be scaled for different throughput requirements by varying the number of slices.

The switch supports multicast efficiently by connecting a single input to multiple outputs. The *Tiny Tera* uses *reverse path self-routing*: each slot, each port transmits a routing tag into the crossbar ahead of any transmitted packet; the routing tag is a 5-bit value that determines the reverse path (it indicates which port will send *to* this port). If forward-path self-routing was used, a 32-bit routing tag would be needed to indicate the set of destinations.

## 2.2 Port

The *Tiny Tera* port is designed to be scalable in data rate and in packet size. The basic switching unit is 64-bits — all packets must be the same length which can be any multiple of 64 bits. The two main challenges in designing a 10Gb/s port interface are: (1) providing memory bandwidth for input and output queues; (2) performing per-packet processing functions, for example, ATM VCI or IP address lookup.

Figure 2 shows how the port architecture is separated into an application-independent packet datapath (Data Slice), and an application-specific port processor (Port Processor).

Packets arrive from the external interface over a set of serial links, each operating at several Gb/s. 64-bits of the packet is cached by each of the Data Slices. We call this 64-bit unit a *chunk*. Each Data Slice forwards the chunk to the Port Processor which decides where in memory the chunk should be stored. The Port Processor sends back a memory address, and a new chunk header. The chunk is updated and written into 64-bit wide SRAM. There are $n$ Data Slices and so the packet is effectively written into a $n$ x 64-bit wide SRAM.

The Port Processor communicates with the scheduler, informing it of newly arrived packets. When the scheduler tells the Port Processor to read from a particular input-queue, the Port Processor issues a read request to the Data Slices, indicating which packet is to be dequeued from memory. The packet is then forwarded over the serial links to the crossbar.

Packets leaving the crossbar are once again buffered by the Data Slices and, under the control of the Port Processor, are stored in output-queues before leaving the system over the external interface. This facility will allow experimentation with switch speedup requirements.

The advantage of maintaining both input and output queues in the same SRAM is that they may be dynamically partitioned, resulting in more efficient usage. This is particularly important, as the switch is to support variable internal speedup, leading to a variable ratio of memory require-ments at input and output. Sharing the SRAM between input and output queues requires two read- and two write operations per packet time; or approximately 10ns per memory operation. This is achievable in commercial memory devices.

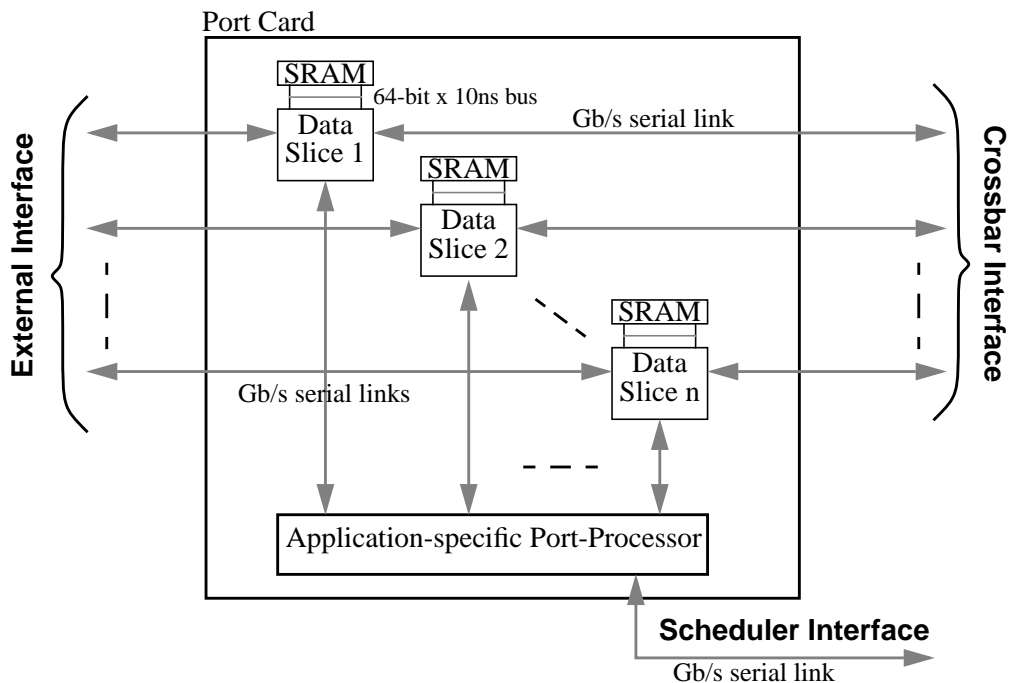

Figure 2 Architecture of port card. The Data Slice components are application-independent, and switch 64-bit *chunks*. The Port-Processor is application-dependent and processes, for example, ATM cell headers or IP addresses.

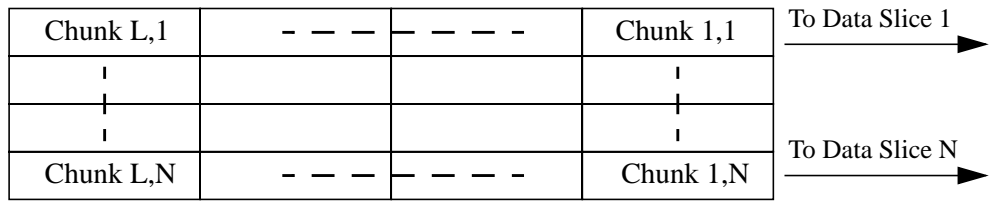

**Figure 3**  A packet can be viewed as a 2-D structure of 64-bit chunks.

The Data Slice is designed to be application-independent. Each Data Slice switches and buffers chunks that are any multiple of 64-bits. The Data Slice is designed so that ports can be built with different data rates by using a different number of slices; each slice will buffer more than one chunk per packet; in fact, a packet can be thought of as a 2-dimensional structure, as shown in Figure 3. A packet consists of multiple chunks processed by multiple Data Slices.

The Port Processor is designed to be application-dependent. For example, in our first implementation, the Port Processor will process 53-byte ATM cells, performing virtual-circuit lookup, and implement the ATM Forum Available Bit Rate (best-effort) Traffic Management standard. A different Port Processor could, for example, implement IP-routing with segmentation and reassembly of packets across the switch core.

# 3  Queueing and Scheduling

The port uses separate structures for buffering unicast and multicast packets; the input-queueing structure is shown in Figure 4.

For unicast packets, the port maintains a separate FIFO queue for each output. This scheme, known as Virtual Output Queueing (VOQ), eliminates HOL blocking because a packet cannot be held up in a queue behind a packet that is destined for a different output. Although slightly more complex (an $N \times N$ switch now maintains $N^2$ input FIFOs), VOQ requires no additional memory bandwidth; at most one packet can arrive and depart from each input per packet-time.

Unfortunately, it is impracticable to eliminate HOL blocking for multicast packets — to do so would require each input to maintain a separate queue for each possible set of destinations. For a 32-port switch this would mean maintaining $(2^{32} - 33) > 4$ billion different queues! In fact, we have found that there is little benefit obtained from maintaining multiple queues unless the number of queues approaches $2^{32}$. Instead, the *Tiny Tera* port maintains only a single FIFO queue for all multicast packets.

## 3.1  Scheduling Unicast Packets

When VOQ is used for unicast packets, the switch requires a scheduling algorithm [5][7] that examines the contents of the $N^2$ input-queues at the beginning of each packet time, deciding which ones will be served. A good scheduling algorithm should be fast, simple, fair, and efficient. (e.g. if the input-lines operate at 10Gb/s, and each packet is a 53-byte ATM cell, the scheduling algorithm must make its decision in less than 42ns).

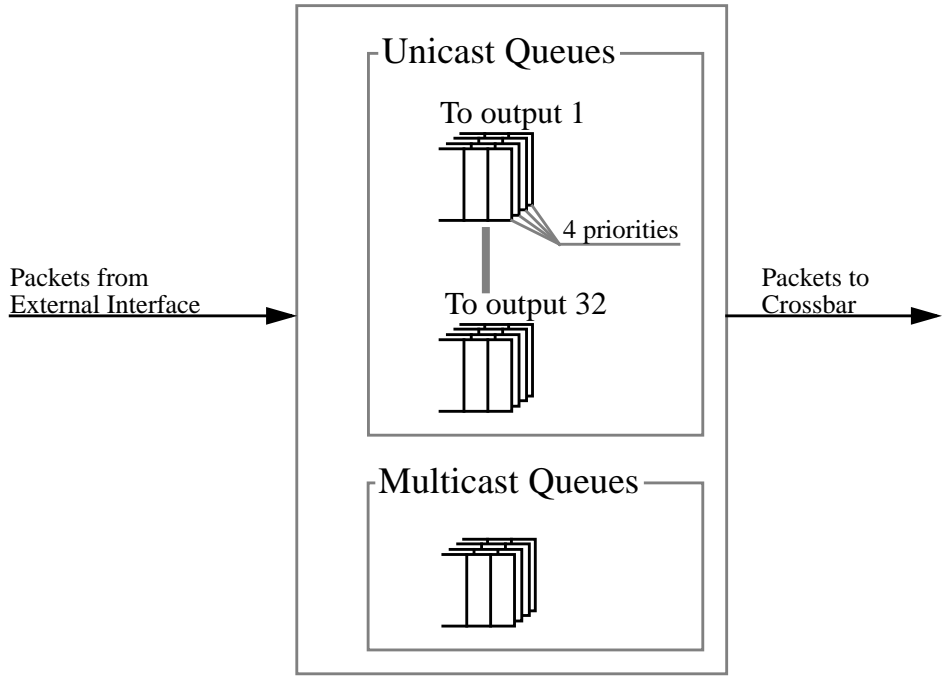

Figure 4   Input-queueing structure at each port.

In previous work, we have shown that the maximum size bipartite graph matching algorithm and the maximum weight bipartite graph matching algorithms (the longest queue first (LQF) and the oldest cell first (OCF)) can achieve 100% throughput [8]. Unfortunately, these algorithms are known to be impractical for implementation in fast and simple hardware, requiring a running time of complexity $O(N^3 \log N)$ [10].

However, we have developed practical, heuristic scheduling algorithms: $i$SLIP, $i$LQF, and $i$OCF [7]. $i$SLIP is an iterative algorithm that provides high efficiency for best-effort traffic and yet is simple to implement in hardware. The algorithm achieves fairness using independent round-robin arbiters at each input and output. Simple round-robin arbiters experience output contention, which limits throughput to just $\left(1 - \frac{1}{e}\right) \approx 63\%$. With a simple modification, $i$SLIP overcomes this problem by causing the arbiters to slip with respect to each other — a match in one slot leads to a larger and faster match in the next slot.

The algorithm behaves as follows. All inputs and outputs are initially unmatched and only those inputs and outputs not matched at the end of one iteration are eligible for matching in the next. Connections made in one iteration are never removed by a later iteration, even if a larger

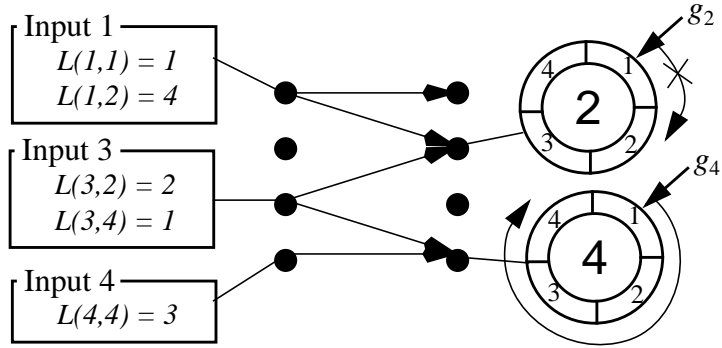

a) Step 1: *Request*. Each input makes a request to each output for which it has a packet.

Step 2: *Grant*. Each output selects the next requesting input at or after the pointer in the round-robin schedule. Arbiters are shown here for outputs 2 and 4. Inputs 1 and 3 both requested output 2. Since $g_2 = 1$ output 2 grants to input 1. Note that pointers $g_1$, $g_2$ and $g_4$ are not updated until Step 3.

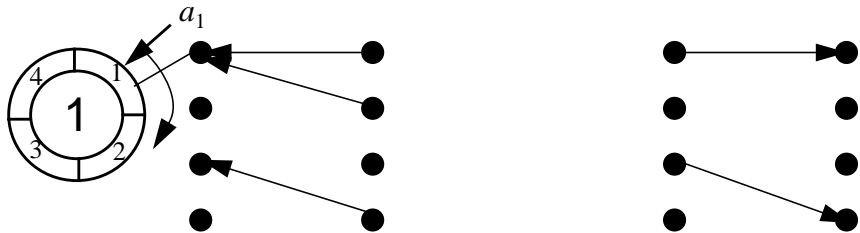

b) Step 3: *Accept*. Each input selects at most one output. The arbiter for input 1 is shown. Since $a_1 = 1$ input 1 accepts output 1. $a_1$ is updated to point to output 2. The pointers at *matched* outputs $g_1$ and $g_4$ are updated. Note that the pointer at the "unsuccessful" output, $g_2$ is not updated.

c) When the arbitration has completed, a matching of size two has been found. Note that this is less than the maximum sized matching of three.

Figure 5  Example of *one iteration* of the *i*SLIP algorithm.

sized match would result. The three steps of each iteration operate in parallel on each output and input and are as follows:

> **Step 1.** *Request*. Each unmatched input sends a request to every output for which it has a queued packet.
>
> **Step 2.** *Grant*. If an unmatched output receives any requests, it chooses the one that appears next in a fixed, round-robin schedule starting from the highest priority element. The output notifies each input whether or not its request was granted. The pointer $g_i$ to the highest priority element of the round-robin schedule is incremented (modulo $N$) to one location beyond the granted input if and only if the grant is accepted in Step 3 *of the first iteration.*
>
> **Step 3.** *Accept*. If an unmatched input receives a grant, it accepts the one that appears next in a fixed, round-robin schedule starting from the highest priority element. The pointer $a_i$ to the highest priority element of the round-robin schedule is incremented

(modulo *N*) to one location beyond the accepted output *only if this input was matched in the first iteration*.

Figure 5 illustrates the three-step arbitration of *i*SLIP for one iteration.

The *i*SLIP algorithm has the following properties:

**Property 1.** For independent arrivals uniformly distributed over all outputs, *i*SLIP achieves 100% throughput with just a single iteration, with more iterations, the queueing delay is reduced. See Figure 6.

**Property 2.** No connection is starved; because of the requirement that pointers are not updated after the first iteration, an output will continue to grant to the highest priority requesting input until it is successful.

**Property 3.** For *i*SLIP with one iteration, and under heavy load, queues with a common output all have the same throughput.

**Property 4.** The algorithm will converge in at most *N* iterations. Simulation suggests that on average, the algorithm converges in fewer than $\log_2 N$ iterations.

**Property 5.** The deterministic nature of *i*SLIP reduces the burstiness of traffic as it transits the switch.

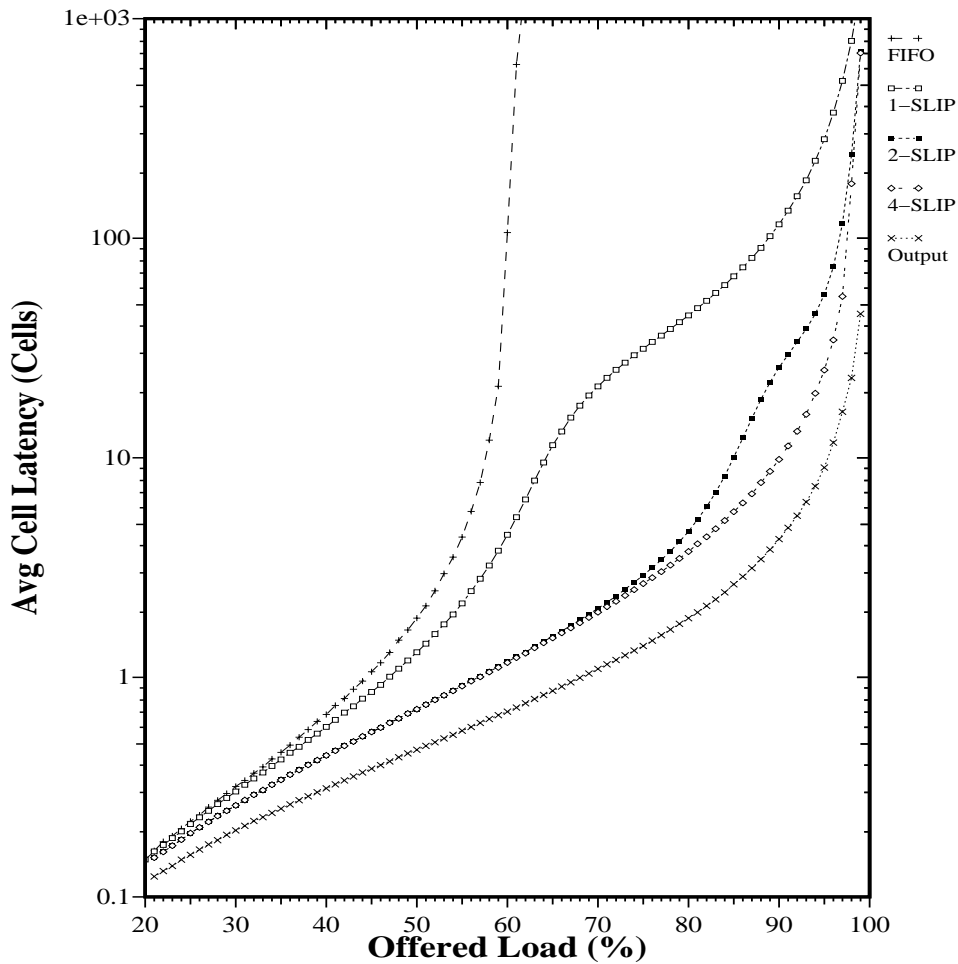

Figure 6  Performance of *i*SLIP for 1,2 and 4 iterations compared with FIFO and output queueing for independent arrivals with destinations uniformly distributed over all outputs. Results obtained using simulation for a 16x16 switch. The graph shows the average delay per packet, measured in slots, between arriving at the input buffers and departing from the switch.

Note that the algorithm will not necessarily converge to a maximum sized match. At best it will find a *maximal* match: the largest size match without removing connections made in earlier iterations. For a more detailed consideration of the performance of *i*SLIP, refer to [7].

Packet delay can be reduced by increasing the number of iterations; via simulation we find that four iterations are sufficient for a 32x32 switch. Perhaps most importantly, we find that a centralized scheduler for a 32x32 switch can be implemented on a single chip, and exploratory designwork suggests that *i*SLIP, implemented in current CMOS technology, can perform one iteration in less than 10ns.

A straightforward implementation of *i*SLIP is shown in Figure 7a; for a 32-port switch the hardware size is dominated by 64 round-robin arbiters. Each arbiter, shown in Figure 7b, consists

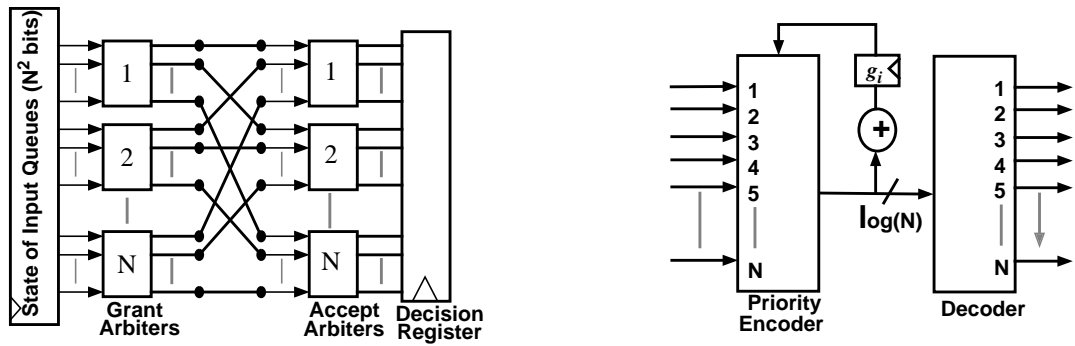

a) Interconnection of 2N arbiters to implement *i*SLIP for an NxN switch.

b) Round-robin *grant* arbiter for *i*SLIP algorithm. The priority encoder has a programmed highest-priority, $g_i$. The *accept* arbiter at the input is identical.

Figure 7   Implementation of *i*SLIP.

of a programmable priority encoder, a register holding a highest-priority pointer, and a decoder. The value of each priority pointer is incremented (updated to one location beyond its matched input or output) only when the matching occurs in step three of the arbitration, otherwise the value remains unchanged.

## 3.2  Scheduling Multicast Traffic

The crossbar performs multicast by simultaneously delivering packets to multiple destinations. There are two service disciplines that can be used. The first is *no fanout-splitting* in which all of the copies of a packet must be sent at the same time. If the packet does not win access to all of the outputs that it desires, it is not copied to any of them, and must try again in the next slot. The second discipline is *fanout-splitting* in which packets may be delivered to output ports over any number of slots. Only those copies that are unsuccessful in one slot continue to contend for output ports in the next slot.

The *Tiny Tera* employs fanout-splitting because it is work-conserving, enabling a higher throughput, yet requiring little increase in implementation complexity. For example, Figure 8 compares the average packet latency (via simulation) with and without fanout-splitting with a random scheduling policy. It is clear that without fanout splitting, performance is seriously degraded. Fortunately fanout-splitting is simple to support; we need only one extra signal from the scheduler to inform each input port when a HOL packet has finished service.

In addition to using fanout-splitting, our scheduling algorithms employ *residue concentration* [9]. To explain residue concentration, we separate packets into input packets and output packets; an input packet is a packet in an input queue, which generates multiple copies; each copy is called an output packet. In implementing fanout-splitting, the scheduler discharges some output packets, possibly leaving behind some residual (unexpanded) output packets in the queue. The residual output packets are called the *residue*. Any work-conserving policy will leave the same residue; it is up to the policy to determine how to distribute the residue over the input ports; the decision on where to place the residue uniquely defines the scheduling policy. Determining the residue is equivalent to determining the transmission schedule.

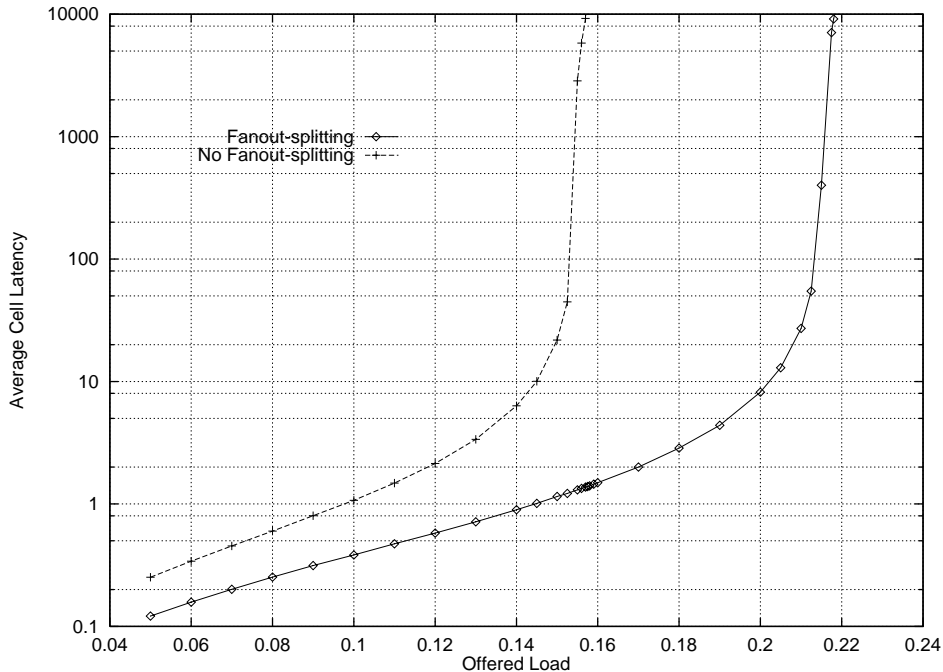

Figure 8  Random Schedule: Average packet latency vs. offered load, with uncorrelated input traffic without destinations uniformly distributed over all outputs, and average fanout of four.

Elsewhere [9], it is argued analytically and via simulation that a policy concentrating the residue on as few inputs as possible will lead to the highest throughput. In particular, for the case of a 2x$N$ switch, it is proved in [9] that, subject to a natural fairness constraint, the optimal algorithm is one that always concentrates the residue on the minimum number of inputs and breaks ties in a round-robin fashion. Concentrating residue on the smallest number of inputs allows more HOL input packets to be completely served, thus bringing more new packets forward to the head of the line. We believe that a residue-concentrating algorithm will also maximize throughput [4] for an $M \times N$ switch.

Unfortunately, the optimal residue-concentrating algorithm is too complex to be practicable. Moreover, a residue-concentrating algorithm can lead to starvation for packets that have a large number of destinations. The *Tiny Tera* will therefore employ one of two practical scheduling algorithms that we are developing. The first is called TATRA which uses the fact that the $M \times N$ scheduling problem can be mapped onto a Tetris-like game. The second is called Weight Based Algorithm (WBA) which is designed to use similar hardware to the *i*SLIP unicast scheduling algorithm. Both algorithms attempt to be fair, yet achieve a high throughput [9].

# 4 High Speed Chip-to-Chip Serial Communication Links

The *Tiny Tera* design poses several chip-to-chip communication challenges. First, the crossbar switch and scheduler chips each need to terminate 32 or more communication links, with each link originating from a different board. Second, the links must achieve very high data rates in a noisy digital environment, as they are integrated on chips with equally high bandwidth memory interfaces and complex logic.

To allow a single chip to terminate many links, we have chosen a serial link architecture that departs from the traditional practice of adjusting the link timing in the receivers. Instead, the *Tiny Tera* links have all the phase adjustment circuitry in the transmitters and receivers at one "smart" end. The other, "dumb" end, does not need to do any phase alignment. By careful arrangement of the links, we ensure that all chips with large fan-in are at the dumb ends of the links, greatly reducing the complexity and power consumption of these chips. Figure 9 shows the link between one of the many data slice chips on a port (smart end) and a crossbar slice chip (dumb end). This same link is reused for almost all IC-IC communications on the *Tiny Tera.*

Operation of high data rate communication links in the low voltage, high edge-rate environment of deep sub-micron CMOS chips requires link circuitry with excellent noise tolerance. In addition, the long feedback path for the phase adjustment of the smart transmitter limits the rate at which phase drift can be tracked, requiring very stable clocks. We improve our voltage noise tolerance by using integrating receivers, while clock stability is achieved with a low phase-noise PLL design that compensates for voltage and temperature variations. Steady and precise phase alignment is accomplished with a digitally controlled clock phase interpolator [12].

## 4.1 Clock Distribution

In a multi-board system such as the *Tiny Tera*, performance is often limited by the distribution of high frequency clocks, and the resultant clock skew and phase noise (jitter). While the *Tiny Tera* design uses a careful clock distribution scheme to minimize skew, we also use an on-chip PLL to multiply up the byte clock that is distributed on board to create the transmit and receive clocks. By performing all high speed board-to-board communication with phase aligning serial links, all clock distribution skews, along with wire and package delays, are offset by the phase adjustment feedback loops.

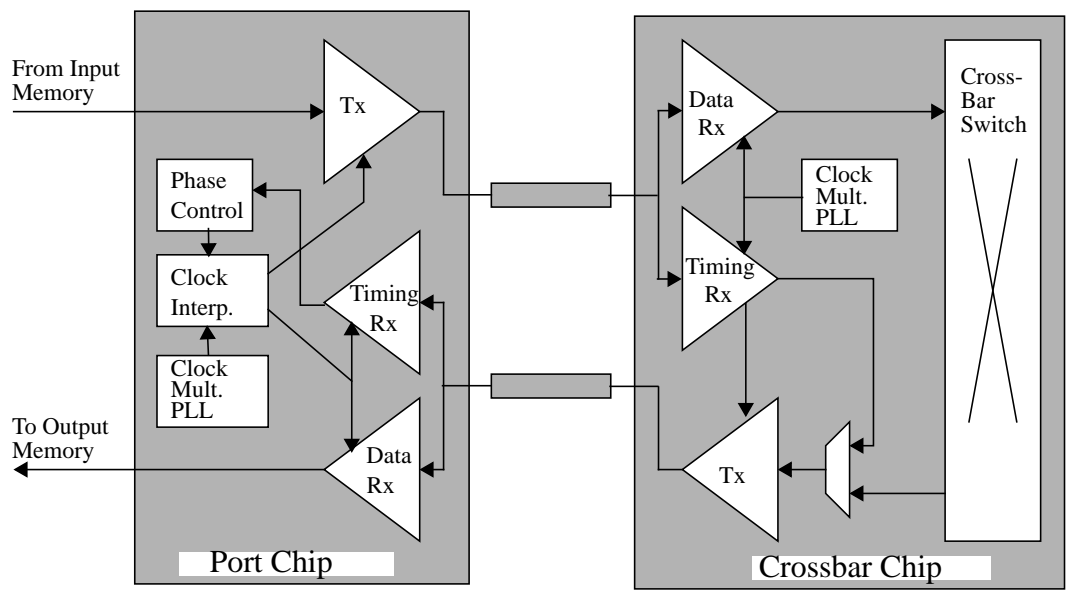

Figure 9   *Tiny Tera* Serial Link Architecture

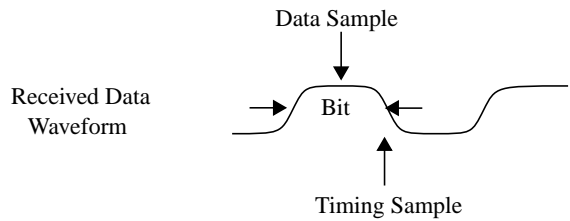

Figure 10  Phase Measurement Waveforms

## 4.2  Phase Control

Phase detection and selection: Two receivers are used in each link. The first is the data receiver which samples the data centered in the bit time, while the second, the timing receiver, samples the data one-half bit time offset, centered about the bit transition. A comparator determines if the timing receiver is sampling before or after the transitions. Phase control logic alters the receive clock to optimize the data receiver sampling window. The timing receiver sampling window is fixed relative to the data receiver window. Figure 10 illustrates a one-to-zero transition, if the timing receiver output is high, the sampling clock was early.

Remote Phase Detection and selection: It is necessary to align the phase of the data from the smart transmitters to the fixed sampling clock of the dumb receivers. The dumb receivers include both a data and a timing receiver, but have no phase adjustment mechanism. The timing information from the dumb receivers is fed back periodically to the smart transmitter, which adjusts the phase of its transmit clock to center the data around the fixed clock in the dumb receiver. Since the transmitter adjusts its phase to match the receiver, all links on the crossbar or scheduler can run off the same clock for transmit and receive with no need for local phase control.

## 4.3  Current Integrating Receivers

Even with proper phase alignment, voltage noise at exactly the sample time (for example switching noise) can still cause errors. Our serial links improve immunity to this by using an input receiver that integrates the value of the input over the entire bit time. This integration makes the timing of any noise irrelevant; the received bit value will depend only on the average value of the signal over the bit time[11].

## 5  Conclusion

We have described the architecture and concepts behind the *Tiny Tera*, a small high performance fixed-size packet switch. The *Tiny Tera* combines the best of switch architecture and scheduling, VLSI datapath, and high performance serial link circuits to create a switching engine that has extremely high performance for a CMOS VLSI-based machine. The *Tiny Tera* is bottlenecked only by memory bandwidth and scheduler speed. We expect it to be a cost effective, compact building block for a wide range of high performance data switching applications pushing the bandwidth available from a single-stage element at least an order of magnitude beyond what is possible today.

# 6 Acknowledgments

We wish to thank Richard Edell of the University of California at Berkeley for discussions with Nick McKeown that lead to the first version of the *Tiny Tera* architecture, as well as Balaji Prabhakar and Ritesh Ahuja who helped in the development of multicast scheduling algorithms. We also thank other members of the *Tiny Tera* team: Ken Chang, Shang-Tse Chuang, Jeff Hsieh, Youngmi Joo, Rolf Muralt, and Brian Stark at Stanford; Ah-Lyan Yee, Helen Chang, P. N.(Ani) Anirudhan, and Sharat Prasad at Texas Instruments.